# Developing a Modular Compiler for a Subset of a C-like Language

Debasish Dutta[1] Neeharika Sonowal[2] and Irani Hazarika[3]

[1,2,3]Department of Computer Science, Gauhati University

**Abstract:** The paper introduces the development of a modular compiler for a subset of a C-like language, which addresses the challenges in constructing a compiler for high-level languages. This modular approach will allow developers to modify a language by adding or removing subsets as required, resulting in a minimal and memory-efficient compiler. The development process is divided into small, incremental steps, where each step yields a fully functioning compiler for an expanding subset of the language. The paper outlines the iterative developmental phase of the compiler, emphasizing progressive enhancements in capabilities and functionality. Adherence to industry best practices of modular design, code reusability, and documentation has enabled the resulting compiler's functional efficiency, maintainability, and extensibility.

The compiler proved to be effective not only in managing the language structure but also in developing optimized code, which demonstrates its practical usability. This was also further assessed using the compiler on a tiny memory-deficient single-board computer, again showing the compiler's efficiency and suitability for resource-constrained devices.

## 1 Introduction

Compilers are software systems that translate human-written source code into machine-understandable code that can be executed. Generally, constructing compilers for complex, high-level languages tends to be excessively intricate[4] , posing a significant challenge for developers aiming to delve into compiler design.

This paper addresses these challenges by proposing the development of a modular compiler through which one can add or remove a subset of the source language per their needs, leading to a minimalistic and efficient compiler. This approach aims to streamline the complexity inherent in compiler construction, enhancing efficiency and reducing barriers for developers seeking to create custom-use compilers. The development of the compiler is divided into small, incremental steps, with each step resulting in a fully functioning compiler for an expanding subset of the language. Notably, the generated assembly code at each step is directly translatable and executable on the target hardware, providing a tangible output that reinforces the learning experience. The resulting compiler exhibits functional efficiency, maintainability, and extensibility by adhering to industry best practices of modular design, code reusability, and documentation. Also, the compiler developed in this research project offers several valuable output capabilities. Alongside the assembly code generation, the compiler can produce detailed token lists, accompanied by the corresponding symbol table, enabling comprehensive analysis of the source code structure. Furthermore, the compiler's ability to generate the Abstract Syntax Tree (AST) provides insights into the hierarchical organization of the code, facilitating advanced optimizations and potential code transformations. By combining a systematic and modular approach to compiler construction, this paper presents a valuable contribution to the field.

## 1.1 Related Works

Ghuloum's paper[3] introduces an incremental approach to designing and implementing a compiler. The paper focuses on the Scheme programming language and the x84–32bit architecture. This process stands out as it allows for the transformation of source code into assembly programs, enabling the generation of a fully functional binary right from the outset. It highlights the ability to generate a working binary directly from the source code, making it a valuable resource for understanding the compiler construction process. Also, in their paper, Dr. Nwanze A and N.N Daniel[2] comprehensively examine Compiler Design, covering all compiler phases and emphasizing the importance of proper design principles. Their research offers valuable insights into the functionalities and challenges of lexical analysis, syntax analysis, semantic analysis, code generation, and optimization. Through an extensive review, the critical research contributions and identified gaps can be addressed in this paper. The insights gained from these studies serve as a foundation for the approach, providing a solid theoretical framework and inspiration for the work. The paper by Abubakar[1] gave an overview of the compiler construction process. It emphasized the role compilers play in translating high-level language into efficient and optimized machine code without altering its original meaning. The paper serves as a comprehensive guide to understanding the intricacies of the compiler construction process.

## 1.2 Problem Definition

The primary objective of this work is to design and implement a compiler in an incremental fashion so that it becomes capable of accepting a significant subset of from a selected source language and generating architecture-specific assembly code.
For this purpose, a subset of the C programming language has been considered the possible input to the compiler. This subset includes essential language constructs, data types, control structures, and function definitions. The work aims to provide a solid foundation for learning and understanding compiler construction by focusing on a limited subset of C.

Again, C was chosen as the implementation language for its performance, efficiency, and extensive libraries that support compiler-related tasks. C provides low-level control, memory management capabilities, and system-level solid programming support, enabling optimal performance and efficient code generation. Its built-in data types, control structures, and operators are well-suited for implementing different compiler stages.

The target architecture, which the compiler will use, consists of an ARM Cortex-A processor, a popular choice for high-performance computing in mobile and embedded systems, and a 1st generation M1 chip. The ARMv8 architecture offers a range of advanced features and instructions that optimize code execution and enhance performance. The Raspberry Pi 3B+, chosen as the target platform to run the compiler, is built on the ARMv8-A architecture. By tailoring the compiler to generate assembly code optimized explicitly for the ARMv8-A architecture, this work aims to produce well-suited code for deployment on the Raspberry Pi 3 and other ARM-based devices.

## 2 Development Phases of the Compiler

As mentioned in section 1.2, this work presents an iterative development methodology for constructing a modular compiler. The proposed methodology consists of a series of phases aimed at progressively enhancing the functionality and capabilities of the compiler, as discussed below-

### 2.1 Phase 1: Design the Base Grammar of the Source Language

This includes a foundational structure, i.e., the main function and syntax for subsequent language constructs.

```
program::= "function" | ε;;
function::= "FUN" IDENTIFIER "(" parameter_list ")" "{" block "}";
parameter_list::= ε;
block ::=<statement><block> |ε;
```

### 2.2 Phase 2: Develop the Base Compiler

Enable the generation of executable binary code from the given base grammar of the source language. The work consists of two parts- the development of the front end of the base compiler and the development of the back end of the base compiler have been shown.

**Front-end of the compiler**.

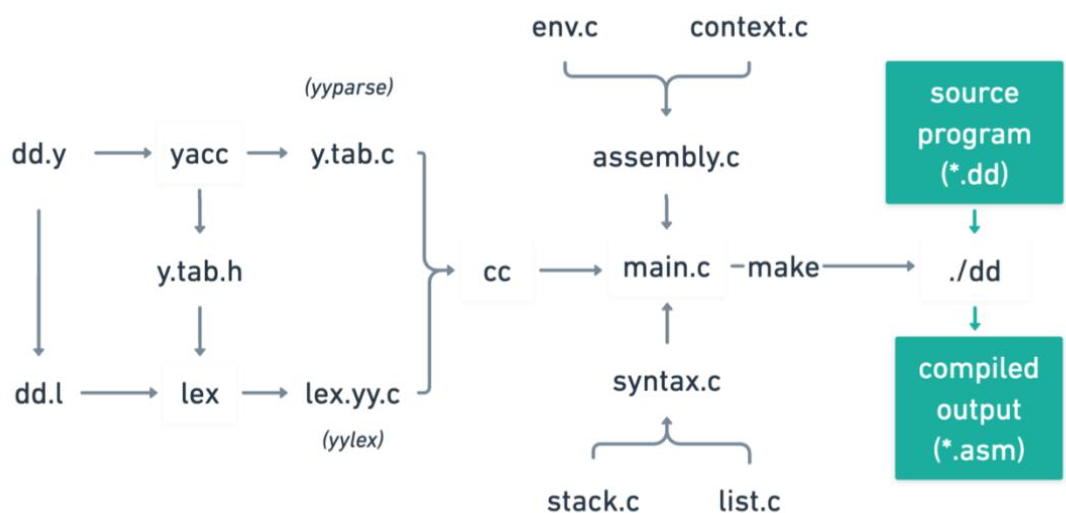

**The back-end of the compiler.**

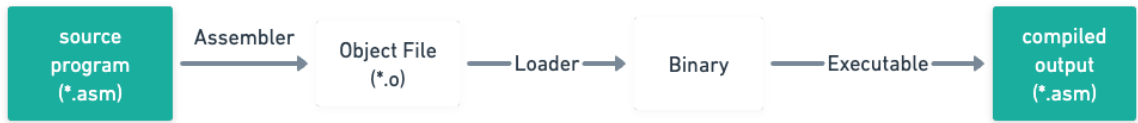

*A. Assembly to Intermediate.*

In the first step, the intermediate assembly code generated by the code generator is assembled into machine code. This conversion process is performed by an assembler, which translates the human-readable assembly code into executable instructions that the target architecture can understand. The result is an object file that contains the machine code and relevant metadata.

*B. Linking to Object Code.*

Once the intermediate assembly code is successfully assembled, the next step is to link the object code. This process combines the object code with necessary linker scripts or runtime libraries to create a complete binary file. The linker resolves memory addresses and symbols, ensuring all dependencies are correctly connected. This results in an object file that can be executed on the target platform.

*C. Executable Binary Generation.*

The final step in the back-end process is generating the executable binary. This is achieved by combining the linked object code with the necessary system libraries and resources. The build system ensures that all the components are appropriately integrated, producing a standalone binary file that can be directly executed on the target architecture.

*D. Building all*

A build system is employed to automate the compilation, assembly, and linking steps to streamline the build process. This build system, consisting of a combination of "Make" and a "zsh" script, orchestrates the various stages of the back end. It ensures that all the required components are built in the correct order, resulting in a fully functional compiler capable of generating the desired executable code.

**Testing the compiler**.

Testing is crucial to compiler development to ensure its correctness and reliability. A testing script, implemented using the "zsh" shell, has been created to automate the execution of test cases. The script covers valid and invalid inputs, testing the compiler's behavior in different scenarios. It prepares the inputs, builds the compiler, executes the test cases, captures the output, and verifies the expected results. After testing, the script performs clean-up tasks to remove temporary and generated files, maintaining a clean working environment.

## 2.3 Phase 3: Expanding the Compiler in Subsets

In this phase, the base compiler's capabilities are extended by incorporating additional language constructs and features. For this purpose, a number of small subsets of the source language are considered and they are added to the compiler incrementally one by one. If *N* numbers of such subsets are added to the compiler, then *N* numbers of stages will be there, where $i^{th}$ stage will add $i^{th}$ subset of the source language to the compiler. In this work, six numbers of subsets have been considered. Hence there will be six stages as described below. For each $i^{th}$ stage the following works have to be done-

*1) Choose a small subset of the source language for added to the compiler*:  Select a targeted subset of the source language to extend the capabilities of the compiler.

*2) Design the grammar for the selected subset of the source language:* Next, the grammar is extended to incorporate the new subset of the language with the previous one to get the complete grammar.

*3) Incorporate the subset into the compiler*: Implement the expansion of the compiler to include the chosen subset, producing assembly output for the expanded language constructs. As the base compiler construction, it also consists of two parts- front end and backend.

*4) Design comprehensive test cases*: Create diverse test cases to evaluate the compiler's functionality, covering both valid and invalid scenarios.

**Stage 1: Add the 1st subset of the source language to the compiler.**

*Choose the 1st language subset: Basic expression evaluation.*
In this 1st stage, a single construct i.e. the return statement is considered as the 1st subset to be added to the compiler. It allows the programmer to specify a value to be returned from a function or program. Here the basic construct is considered to establish the foundation of the compiler and ensure that it can handle the fundamental structure of the language.

*Design the grammar for the 1st language subset.*
The language grammar is designed to recognize and parse return statements. It defines the syntax and structure of a valid return statement within the subset.

```
<block>: := <statement><block> |ε;

<statement> : := 'RETURN'<expression>';'
                        | <TYPE><IDENTIFIER>'='<expression>';'
                        | <expression>';'
```

```
<expression> : := <NUMBER>
                          | <IDENTIFIER>
```

*Incorporate the 1st language subset into the compiler.*
The compiler is implemented to handle the subset by developing the necessary components. This includes the scanner, responsible for tokenizing the source code, the parser, which constructs the abstract syntax tree (AST) based on the grammar rules, and the code generator, which generates the corresponding assembly instructions for the return statements.

    a.   **Scanner**
    The scanner is implemented to recognize tokens such as identifiers, numbers, operators, and return expressions.
    b.   **Parser**
    Define grammar rules for return expressions and modify the parser to handle return statements in the language subset.
    c.   **Code Generation**
    Extend the code generator to handle return expressions by mapping them to the corresponding assembly or machine code instructions.

```
emit_instr_format(out, "mov",
"X0, #%d", syntax->immediate->value);
```

*Design comprehensive test cases.*
Test cases are created to verify the functionality of the compiler for valid and invalid input scenarios. Valid test cases include various return statements, ensuring that the compiler correctly handles different types of values. For example-

```
        fun main() {
            return 9;
         }
```

In this valid test case, a function named main is defined, which returns the integer value 100. This test case verifies that the compiler can correctly handle a valid function definition and return statement.

Invalid test cases cover scenarios such as missing return statements or incorrect syntax. For example-

```
         fun main( {
            return 0;
         }
```

In this invalid test case, the function definition for main is missing a closing parenthesis ')'. This test case aims to verify the compiler's ability to detect and report syntax errors, specifically when encountering missing or mismatched parentheses.

**Stage 2: Add the 2nd subset of the source language to the compiler.**
In this stage, second subset of the source language has been added to the compiler as given below-

*Choose the 2nd language subset: Unary Operators.*
Building upon the foundation established in Stage 1, Stage 2 introduces unary operators to the language subset. Unary operators are operators that act on a single operand and modify its value. The chosen subset of unary operators includes bitwise negation (~), logical negation (!), and negation (-).

*Design the grammar of the 2nd language subset.*
The language grammar is extended to incorporate unary operators into the expression block. The grammar rules are modified to handle unary operator expressions.

```
<expression> ::=<NUMBER>
               | <IDENTIFIER>
               | <IDENTIFIER>'='<expression>
               | '-'<expression>
               | '~'<expression>
               | '!'<expression>
```

*Incorporate the 2nd language subset into the compiler.*
The scanner, parser, and code generation modules are updated to recognize and handle unary operator expressions. The scanner identifies unary operators as tokens, the parser constructs the AST accordingly, and the code generator generates appropriate assembly instructions for the unary operator expressions.

  A. **Scanner**
     Modify the scanner to recognize unary operators and updating the tokenization process accordingly.
  B. **Parser**
     Adding grammar rules to handle unary expressions and enhance the parsing logic to construct the AST for unary expressions.
  C. **Code Generation**
     Extend the code generator to generate the appropriate assembly or machine code for unary expressions, considering the specific unary operators defined in the language subset.

```
emit_instr(out, "neg", "X0, X0");
```

By successfully completing Stage 2, the compiler gains the capability to handle unary operators, enhancing the expressiveness and functionality of the custom language. This incremental approach ensures that each stage builds upon the previous one, creating a solid foundation for subsequent stages while ensuring that existing functionality remains intact.

*Design comprehensive test cases.*
Test cases are designed to cover different combinations of unary operators and operands. These test cases validate the correct parsing and code generation for unary operator expressions. For example-

```
fun main() {
    return !7;
}
```

In this above valid test case, a function named main is defined, which returns the logical negation of the decimal value 7. This test case verifies that the compiler can correctly handle a valid function definition and return statement.

```
fun main() {
    return +6
}
```

In this above invalid test case, the function definition for main is using a unknown operator'+'. This test case aims to verify the compiler's ability to detect and report syntax errors, specifically when encountering mismatched operators.

**Stage 3: Add the 3rd subset of the source language to the compiler.**
In this stage, third subset of the source language has been added to the compiler as given below-

*Choose the 3rd language subset: Primary Binary Operators.*
In Stage 3 of the compiler development, the focus is on introducing primary binary operators to the language subset. Primary binary operators are operators that operate on two operands and perform basic arithmetic operations. The chosen subset of primary binary operators includes addition (+), subtraction (-), and multiplication (*). These operators allow the programmer to perform basic arithmetic calculations on two values.

*Design the grammar of the 3rd language subset.*
The language grammar is expanded to incorporate primary binary operators into the expression block. The grammar rules are modified to handle binary operator expressions involving addition, subtraction, and multiplication.

```
<expression> ::= ...
        | <expression>'+'<expression>
        | <expression>'-'<expression>
        | <expression>'*'<expression>
```

*Incorporate the 3rd language subset into the compiler.*
The scanner, parser, and code generation modules are updated to recognize and handle primary binary operator expressions. The scanner identifies binary operators as tokens, the parser constructs the AST accordingly, and the code generator generates appropriate assembly instructions for the binary operator expressions.

- **A. Scanner**
  Update the scanner to recognize binary operators and ensuring they are tokenized correctly.
- **B. Parser**
  Enhancing the parsing to handle binary expressions, including the precedence and associativity of binary operators.
- **C. Code Generation**

Extending the code generator to generate the appropriate assembly or machine code for binary expressions, considering the specific binary operators defined in the language subset. This is done in such a way that the assembly should be able to handle nested operations also.

```
emit_instr_format(out, "ldr",
  "X1, [sp, #%d]" , stack_offset);
emit_instr_format(out, "add",
  "X0, X0, X1");
```

*Design comprehensive test cases.*
A set of test cases is designed to cover various combinations of primary binary operators and operands. These test cases ensure that the compiler correctly parses and generates code for binary operator expressions. For example-

```
fun main() {
    return 10 + 3;
}
```

In this valid test case, a function named main is defined, which returns the difference of 10 & 3. This test case verifies that the compiler can correctly handle a valid function definition and return statement.

```
fun main() {
    return 2 -+ 3 ;
}
```

In this invalid test case, the function definition for main has a malformed operator '-+'. This test case aims to verify the compiler's ability to detect and report syntax errors.

**Stage 4: Add the 4th subset of the source language to the compiler.**

In this stage, fourth subset of the source language has been added to the compiler as given below-

*Choose the 4th language subset: Additional Binary Operators.*
Stage 4 focuses on introducing additional binary operators to the language subset, expanding its expressiveness and functionality. The chosen subset of additional binary operators includes logical operators such as logical AND (&&), logical OR (||), and equality operator (==). These operators allow the programmer to perform logical operations and comparisons between two values. Also are chosen the comparison operators that are GREATER THAN ('>'), LESS THAN ('<'), GREATER EQUALS ('>=') and LESS EQUALS ('<=').

*Design the grammar of the 4th language subset.*
The language grammar is extended to incorporate additional binary operators into the expression block. The grammar rules are modified to handle binary operator expressions involving logical AND, logical OR, and equality comparisons.

```
<expression> ::= ...
            | <expression>'AD'<expression>
            | <expression>'ORR'<expression>
            | <expression>'EQ'<expression>
            | <expression>'GRT'<expression>
            | <expression>'LST'<expression>
            | <expression>'GT_EQ'<expression>
            | <expression>'LT_EQ'<expression>
```

*Incorporate the 4th language subset into the compiler.*
The scanner, parser, and code generation modules are updated to recognize and handle additional binary operator expressions. The scanner identifies the binary operators as tokens, the parser constructs the AST accordingly, and the code generator generates appropriate assembly instructions for the additional binary operator expressions.

A. **Scanner**
Update the scanner to recognize additional binary operators such as logical operators ('&&', '||') and comparison operators ('>', '<', '==').

B. **Parser**
Ensure proper tokenization of the new binary operators and distinguishing them from other tokens in the input.

C. **Code Generation**
Extending the code generator to generate the appropriate assembly or machine code for binary expressions, considering the specific binary operators defined in the language subset.

```
emit_instr_format(out, "ldr",
    "X1, [sp, #%d]" , stack_offset);
emit_instr_format(out, "cmp", " X1,
```

```
                         X0", stack_offset);
        emit_instr_format(out, "cset", " X0,
ge", stack_offset);
```

*Design comprehensive test cases.*
Test cases are designed to cover various scenarios involving the additional binary operators. These test cases verify that the compiler correctly parses and generates code for logical and equality expressions. For example-

```
fun main() {
    return 10 && 10;
}
```

In this valid test case, a function named main is defined, which returns the value of 10 and 10. This test case verifies that the compiler can correctly handle a valid function definition and return statement. For example-

```
 fun main() {
   return 7 9;
}
```

In this invalid test case, the function definition for main is missing the operator. This test case aims to verify the compiler's ability to detect and report syntax errors, specifically when encountering missing or mismatched operands and operators.

**Stage 5: Add the 5th subset of the source language to the compiler.**
In this stage, fifth subset of the source language has been added to the compiler as

*Choose the 5th language subset: Variables.*
Stage 5 of the compiler development focuses on introducing variable declarations into the language subset. Variable declarations allow programmers to define named storage locations to store data during program execution.

*Design the grammar of the 5th language subset.*
The language grammar is updated to include syntax rules for variable declarations. This includes specifying the keyword for declaring variables, the variable name, and optionally the variable type.

```
<expression> ::= ...
              | <IDENTIFIER>'('<argument_list>')'

<argument_list> ::= <nonempty_argument_list>
              | ε;
<nonempty_argument_list> ::= <expression>','
```

```
<nonempty_argument_list>
                         | <expression>
```

*Incorporate the 5th language subset into the compiler.*
The scanner and parser modules are updated to recognize variable declaration statements and assignment expressions. The code generator is modified to generate assembly instructions for allocating memory for variables and storing values in the assigned memory locations.

### A. Scanner
Enhance the scanner to recognize variable names and keywords related to variable declarations namely 'var'.

### B. Parser
Extend the tokenization process to properly identify and differentiate variables and keywords from other tokens.

### C. Code Generation
Extending the code generator to generate the appropriate assembly code. The code generator is modified to handle variable declarations by reserving memory space and assigning appropriate memory addresses to the declared variables. For this purpose, the variables are stored in separate memory registers and a stack called environment is used which can also be said to be the scope of the variables. The compiler is also enhanced to handle variable assignment statements. This involves parsing assignment expressions, retrieving the value to be assigned, and storing it in the designated memory location for the variable.

```
environment_set_offset(ctx->env,
define_var_statement->var_name,
stack_offset);
ctx->stack_offset -= WORD_SIZE;
write_syntax(out, define_var_statement->init_value, ctx);
emit_instr_format(out, "str", "X0,
          [sp, #%d]\n", stack_offset);
```

*Design comprehensive test cases.*
Test cases are designed to cover various scenarios involving variable declarations and assignments. These test cases verify that the compiler correctly handles variable declaration syntax, memory allocation, and assignment operations. For example-

```
fun main() {
    var a = 9;
```

```
        return a;
    }
```

This valid test case is designed to evaluate the compiler's ability to handle variable declarations and assignments, as well as returning a variable's value. The expected behaviour is that the compiler correctly generates an Abstract Syntax Tree (AST) representing the program's structure, identifies the tokens in the source code, and produces the corresponding assembly code.

```
    fun main() {
        a = 8;
        return 0;
    }
```

This invalid test case is designed to test the error handling of variable declaration. Here the variable 'a' is assigned without prior declaration. The expected behaviour is that the compiler detects the error and provides a meaningful error message indicating the undeclared variable.

**Stage 6: Add the 6th subset of the source language to the compiler.**
In this stage, sixth subset of the source language has been added to the compiler as given below-

*Choose the 6th language subset: Conditional statements.*
Stage 6 of the compiler development focuses on introducing conditional statements, specifically if-else constructs, into the language subset. Conditional statements allow programmers to control the flow of execution based on specified conditions.

*Design the grammar of the 6th language subset.*
The language grammar is updated to include syntax rules for if-else constructs. This includes specifying the keywords for conditional statements, the condition expression, and the blocks of code to be executed based on the condition.

```
    <statement> ::= ...
                | IF '('<expression>')'
    <block> (ELSE <block>)?
```

*Incorporate the 6th language subset into the compiler.*
1. **Scanner**
   Update the scanner to recognize conditional keywords 'if' and 'else'.
2. **Parser**
   Ensuring proper tokenization of the conditional keywords and distinguish them from other tokens in the input.

### 3. Code Generation

During code generation, it's important to consider the control flow of the program and it ensure that the generated assembly code correctly handles the execution paths based on the conditions. The compiler needs to evaluate the condition expression in if statement and determine whether to execute the corresponding block of code or the optional else block. This involves generating assembly code that performs the necessary comparisons and conditional branching.Also, variable scoping and memory management should be taken into account to avoid conflicts and ensure proper storage and retrieval of variables within the conditional blocks.

```
write_syntax(out,
if_statement->condition, ctx);
char *label_end = fresh_local_label
            ("if_end", ctx);
char *label_else = fresh_local_label
            ("if_else", ctx);
emit_instr(out, "cmp", "x0, #0");
emit_instr_format(out, "beq",
              "%s", label_else);
write_syntax(out,if_statement->then_stmts,
ctx);
emit_instr_format(out,"b","%s",label_end);
emit_label(out, label_else);
if (if_statement->else_stmts != NULL) {
write_syntax(out,
if_statement->else_stmts, ctx);
    }
emit_label(out, label_end);
```

*Design comprehensive test cases.*
Test cases are designed to cover different scenarios involving if-else constructs. These test cases verify that the compiler correctly handles conditional statements, evaluates the condition expression, and executes the appropriate code block based on the condition's outcome. For example-

```
fun main() {
      var a = 9;
      if (a<0){
          return a;
      } else {
          return a-3;
      }
   }
```

This valid test case is designed to evaluate the compiler's ability to handle if-else constructs. The expected behavior is that the compiler correctly generates the AST, evaluates the condition expression (a < 0), and generates assembly code that branches to the appropriate code block based on the condition's outcome.

```
fun main() {
    if (a==9){
        return 1;
    }
}
```

This invalid test case is designed to test the error handling capabilities of the compiler. The variable 'a' is referenced in the condition expression without prior declaration. The expected behavior is that the compiler detects the error and provides a meaningful error message indicating the undeclared variable.

## 3   Results and Discussions

The compiler is capable of producing three intermediate outputs along with the compiled binary per source program, identifiable by the flag provided during compilation. It produces the tokens used in by the scanner, the AST generated by the parser and the assembly code used in the intermediate code generation phase, all before producing the main compiled output. This versatility in output options enable comprehensive analysis and understanding of the compilation process. The availability of token information allows developers to examine the lexical structure of the source code aiding in identifying any potential bugs or inconsistencies. The AST provides a hierarchical representation of the program's syntactic structure, facilitating further analysis and optimization techniques. This detailed view of the program's structure aids in understanding complex code relationships and optimizing code execution. Additionally, the compiler generates assembly code as part of the intermediate code generation phase. This assembly code is an intermediary representation between the high-level source and target machine codes. It provides a human-readable representation of the program's logic and operations, allowing developers to inspect the translation process and make any necessary optimizations or adjustments.

The ability to access these intermediate outputs enhances the learning experience for developers seeking to understand the inner workings of the compiler but also contributes to the overall development process. By examining the tokens, AST, and assembly code, developers gain insights into the compiler's interpretation of their source code and can make informed decisions on code improvements, performance optimizations, and potential language enhancements.

Moreover, the compiler's capability to provide multiple outputs aligns with industry best practices and supports the principle of transparency and understanding in the development process. It promotes collaboration and knowledge sharing among developers, educators, and researchers, fostering a deeper understanding of compiler construction and its implications in real-world applications.

To illustrate the outputs provided by the compiler, the source code written in the high-level language supported by the compiler yields the following outputs:

## 3.1 Input to the compiler

This is an example source code for which the compiler is able to work on. Here, the return statement is completed on stage 1(section 4.3) while the var one in stage 5 (section 4.3) and the function declaration is done in the base compiler stage (section 4.2).

```
fun main() {
   var a = 0;
   a = 9;
   return a;
}
```

## 3.2 Compilations Output

**Front-End Part.**

*Scanner.*
The compiler generates a list of tokens during the scanning phase, representing the source code's lexical elements. These tokens serve as the fundamental building blocks for the subsequent stages of compilation.

```
Tokens
   fun             Token: 258
   main            Token: 260
   (               Token: 263
   )               Token: 264
   {               Token: 265
   var             Token: 259
   a               Token: 260
   =               Token: 274
   0               Token: 262
   ;               Token: 272
   a               Token: 260
   =               Token: 274
   9               Token: 262
   ;               Token: 272
   return          Token: 261
   a               Token: 260
   ;               Token: 272
   }               Token: 266

   Total tokens: 18
```

Here, the token number represents that specific token which is used throughout the compiler to recognize the tokens.

*Parser.*

The following AST generated by the parser represents the hierarchical structure of the source code in program above.

```
---AST---
    TOP LEVEL
        FUNCTION 'main'
            BLOCK
                DEFINE VARIABLE 'a'
                'a' INITIAL VALUE
                    IMMEDIATE 0
                ASSIGNMENT 'a'
                    IMMEDIATE 9
                RETURN
                    VARIABLE 'a'
```

*Code Generator.*
This assembly code below represents the translation of the high-level code defined in above into low-level instructions that can be executed by the target architecture.

```
.global _main
      .align 4

      _main:
          mov       X0, #0
          str       X0, [sp, #-16]

          mov       X0, #9
          str       X0, [sp, #-16]
ldr       X0, [sp, #-16]

          mov       x16, #1;
          svc       #0xFFFF;
```

**Back-end Part.**

*Execution of the Compiled binary.*
The final output of the compiler is a binary file that contains the compiled code corresponding to the source program in above. When executed on the target platform, the binary file exhibits the desired behavior specified by the source code.

```
> ./main
> echo $?
      9
```

## 4 Real World Use Case

In addition to examining the outputs and functionality of the compiler, it is crucial to assess its applicability in real-world scenarios. Thus, this paper uses this compiler on a real world, resource constraint device i.e. the Raspberry Pi 3B+ which has 1gb of memory along with 32gb microSD card space.
The aim is to evaluate the compiled code's execution on this platform to assess its efficiency and suitability for memory-constrained embedded systems. For this purpose, a sample source code is selected from each of the compiler stages and the CPU usage(CPU), memory usage(MU) and execution time(ET) as well as the source file(SF), compiled assembly file(AF) and output binary file(BF) sizes, are measured. It is shown in the table 1.

```
raspberrypi% ./build.sh main.dd
Compiling main...
Building main
fun main() {
    var a=0;
    a = 6;
    return a;
}

Written build/main.asm.
linking and loading with Pi
Running binary main
6
raspberrypi%
```

**Fig. 1.** A snapshot of the compiler running in the Raspberry Pi

**Table 1.** Performance Evaluation Results on the Raspberry Pi

| Program  | CPU (U%) | MU (MB) | ET (ms) | SF (b) | AF (b) | BF (b) |
|----------|----------|---------|---------|--------|--------|--------|
| *return_9* | 1.02   | 1.02    | 102     | 5      | 105    | 856    |
| *not_7*    | 2.00   | 1.48    | 142     | 5      | 151    | 872    |
| *Sub_10_3* | 5.40   | 2.19    | 126     | 5      | 214    | 1680   |
| *and_10*   | 2.25   | 2.08    | 111     | 5      | 215    | 1680   |
| *var_a*    | 6.90   | 2.85    | 230     | 43     | 166    | 1648   |
| *if_else*  | 14.20  | 2.20    | 350     | 100    | 557    | 1824   |

## 5 Conclusion

The compiler developed using this incremental compiler construction process enables the translation of high-level language source code into an executable one with a modular development process. After thorough testing, the compiler has proven highly effective in managing diverse language structures and producing optimized code. The practical implications of the compiler are evident in its real-world use case scenario on the Raspberry Pi platform. By using the modular process and making gradual improvements, it will be helpful to developers and users alike to use develop a compiler based on their specific usage and not thus minimizing their resource usage.

**References**

1. Abubakar, B. S., Ahmad, A., Aliyu, M. M., Ahmad, M. M., & Uba, H. U. (2021). An Overview of Compiler Construction. *International Research Journal of Engineering and Technology (IRJET),* 8(03), 587.
2. Ashioba, D. N., & Daniel, N. N. (2021). *Compiler Construction Detail Design*, 12(8).
3. Ghuloum, A. (2006). An incremental approach to compiler construction. *In Proceedings of the 2006 Scheme and Functional Programming Workshop* (pp. 27–37). Scheme.
4. Grune, D., Van Reeuwijk, K., Bal, H. E., Jacobs, C. J., & Langendoen, K. (2012). Modern compiler design. *Springer Science & Business Media*.
5. Lesk, M. E., & Schmidt, E. E. (1990). Lex—a lexical analyzer generator.
   Johnson, S. C., & Sethi, R. (Eds.). (1979). Yacc: A Parser Generator. In UNIX Vol. II: Research System (10th Ed.). W. B. Saunders Company.
6. Patrignani, M., Devriese, D., & Piessens, F. (2016). On Modular and Fully-Abstract Compilation. *2016 IEEE 29th Computer Security Foundations Symposium (CSF)*, 17–30.
7. Siek, J. G. (2023). Essentials of Compilation: An Incremental Approach in Racket. MIT Press.